\documentclass[aps,preprint]{revtex4_noab}
\usepackage{epsfig}
\usepackage{color}

\def\Bbf{\mathbf B}

\def\ebf{\mathbf e}

\newcommand{\pop}{{\it Phys. Plasmas}}
\newcommand{\jgr}{{\it J. Geophys. Res.}}
\newcommand{\grl}{{\it Geophys. Res. Lett.}}

\renewcommand{\prl}{{\it Phys. Rev. Lett.}}

\renewcommand{\jcp}{{\it J. Comput.Phys.}}

\begin{document}
\title{
Three-Dimensional Geometry of Magnetic Reconnection Induced by Ballooning Instability in a Generalized Harris Sheet
}
\author{
Ping Zhu$^{1,2}$, Amitava Bhattacharjee$^3$, Arash Sangari$^2$, Zechen Wang$^1$, and Phillip Bonofiglo$^2$ 
\\ $^1${\it CAS Key Laboratory of Geospace Environment and Department of Modern Physics \\ University of Science and Technology of China \\ Hefei, Anhui 230026, PRC} 
\\ $^2${\it Department of Engineering Physics \\
University of Wisconsin-Madison \\
Madison, WI 53706, USA}
\\ $^3${\it Princeton Plasma Physics Laboratory \\
Princeton University \\
Princeton, NJ 08543, USA}
}
\begin{abstract}
We report for the first time the intrinsically three-dimensional (3D) geometry of the magnetic reconnection process induced by ballooning instability in a generalized Harris sheet. The spatial distribution and structure of the quasi-separatrix layers, as well as their temporal emergence and evolution, indicate that the associated magnetic reconnection can only occur in a 3D geometry, which is irreducible to that of any two-dimensional reconnection process. Such a finding provides a new perspective to the long-standing controversy over the substorm onset problem, and elucidates the combined roles of reconnection and ballooning instabilities. It also connects to the universal presence of 3D reconnection processes previously discovered in various natural and laboratory plasmas.
\end{abstract}


\maketitle

Magnetic reconnection is believed to play a critical role in the eruptive energy conversion processes that are ubiquitous in laboratory and natural plasmas. Traditionally magnetic reconnection has been often interpreted using 2-dimensional (2D) models, such as those by Sweet-Parker and Petschek~\cite{sweet58a,sweet58b,parker63a,petschek64a}, even though in reality magnetic reconnection always takes place in 3-dimensional (3D) space (e.g.~\cite{pontin11a}). The 2D reconnection models essentially assume or imply the invariance of the magnetic reconnection process in the third spatial dimension perpendicular to the reconnecting field lines. While such 2D representations have been useful in interpreting and understanding many phenomena in 3D, they have also led to confusion and controversy, perhaps exemplified best by the problem of substorm onset in the near-Earth magnetotail. Even after the observations from the highly successful Themis mission are in, there continues to be debate regarding the precise mechanism of substorm onset. While there appears to be compelling evidence for the role of reconnection~\citep{angelopoulos08c}, there is an almost equally plausible case to be made for the ballooning instability~\citep{panov12a}. This controversy is rooted in part in the vexing history of magnetic reconnection in the magnetotail, where it is widely accepted that at near-Earth distances, the tearing mode tends to be strongly suppressed due to the presence of a non-vanishing $B_z$-field, which destroys the X-line of the classic reconnection geometry in 2D.

In this paper, we demonstrate that a novel perspective that reconciles the two conflicting viewpoints regarding substorm onset is one that replaces the notion of an X-line in 2D with that of a Quasi-separatrix Layer (QSL) in 3D. The QSL concept has found extensive use in solar and more recently in laboratory plasma physics, where the presence of boundary conditions such as line-tying often rules out the occurrence of standard separators such as closed magnetic field lines or null-null lines in 3D. When the concept is used here in the context of the Earth's magnetotail for the first time, we find that it provides a natural way to reconcile conflicting points of view regarding the problem of substorm onset.

\begin{figure}
\includegraphics[width=0.49\textwidth]{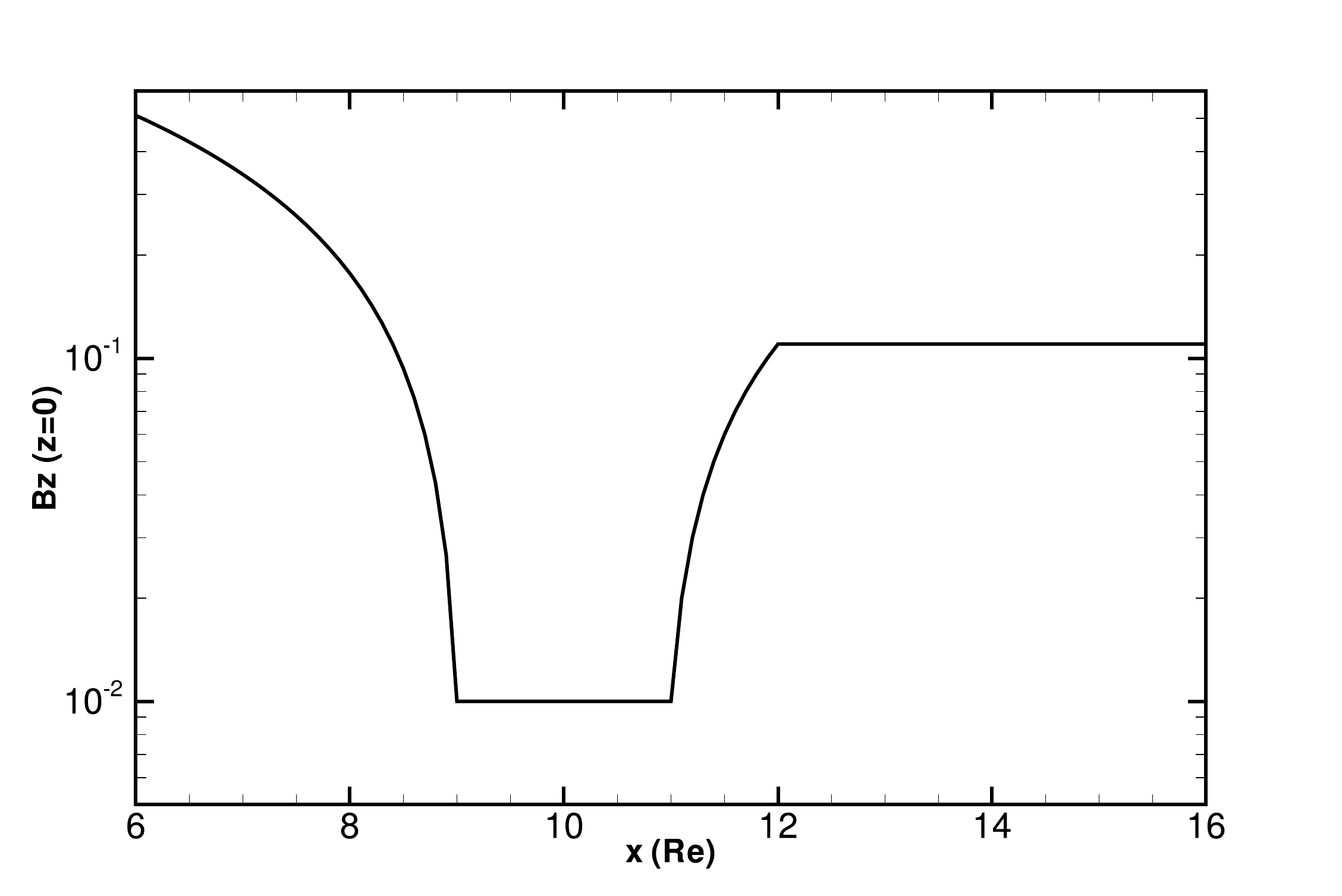}
\includegraphics[width=0.49\textwidth]{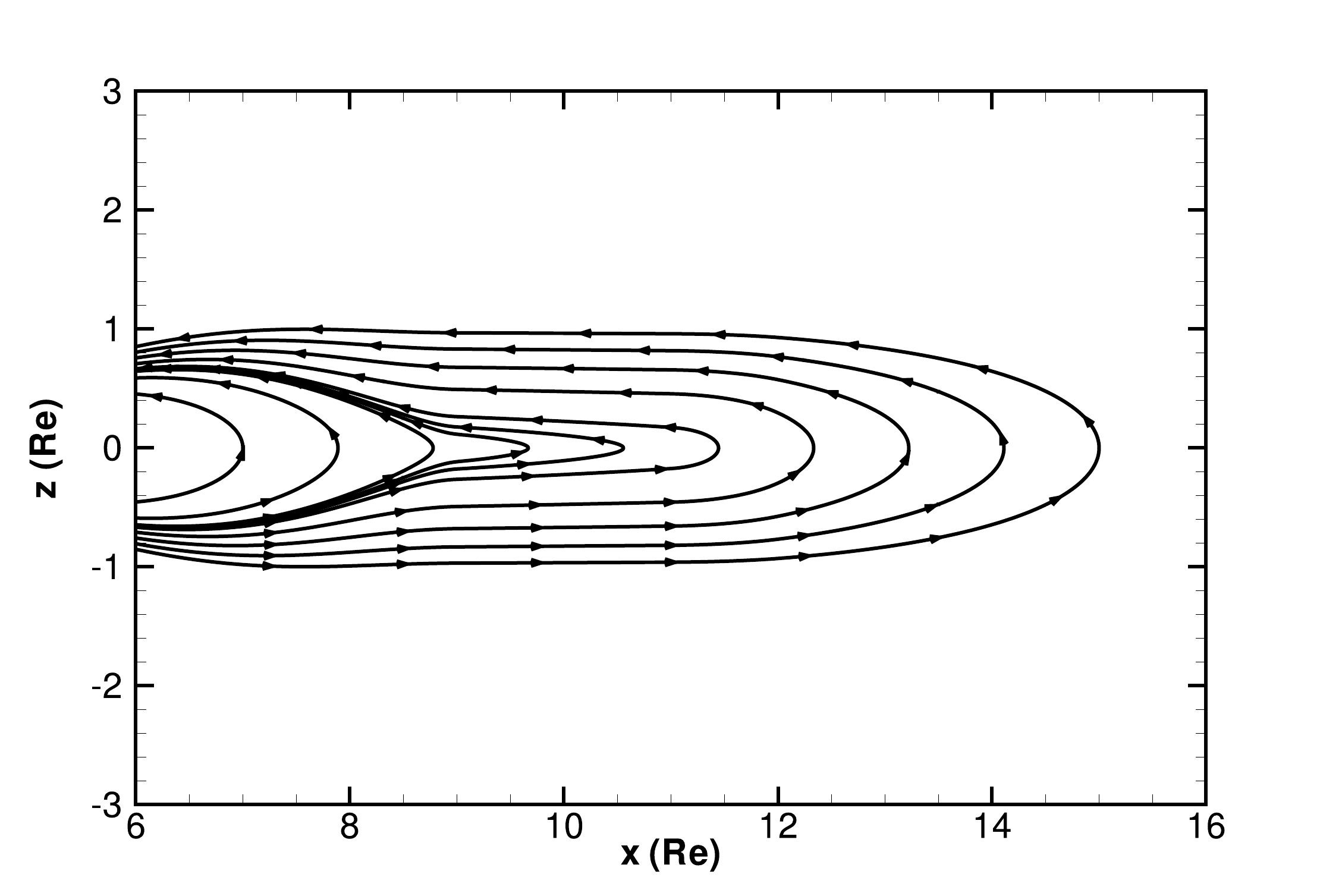}
\caption{$B_z(x,z=0)$ profile (left) and magnetic field lines (right) of a generalized Harris sheet.}
\label{fig:gharris}
\end{figure}

We begin with some background. Our recent simulations based on the resistive single-MHD model implemented in the NIMROD code~\citep{sovinec04a} have found a plasmoid formation process in the generalized Harris sheet that is often used as a proxy to the configuration of the near-Earth magnetotail prior to a substorm onset~\cite{zhu13d,zhu14a}, where the magnetic configuration can be defined in a Cartesian coordinate system as $\Bbf_0(x,z)=\ebf_y\times\nabla\Psi(x,z)$, $\Psi(x,z)=-\lambda\ln{\frac{\displaystyle\cosh{\left[F(x)\frac{z}{\lambda}\right]}}{\displaystyle F(x)}}$, $\ln{F(x)}=-\int B_{0z}(x,0)dx/\lambda$, and $\lambda$ is the characteristic width of the current sheet. The conventional Harris sheet is recovered when $F(x)=1$. The configuration can be further specified with a particular $B_z$ profile that features a minimum
region along the $x$ axis, which could correspond to an embedded thin current sheet (Fig.~\ref{fig:gharris}), such as those often found in global MHD simulations and inferred from satellite observations in the near-Earth magnetotail. Those simulations demonstrate that the embedded thin current is unstable to ballooning mode perturbations, and the nonlinear development of the ballooning instability is able to induce the onset of reconnection and the formation of plasmoids in the current sheet where there is no pre-existing X-point or X-line. The plasmoid instability in our study, which are in the high-S regime ($S>10^4$), should be contrasted with earlier lower-S studies~\citep{birn81a,hesse91a,schindler07a} in which the magnetotail is found to be linearly unstable to resistive tearing modes. In contrast, the generalized Harris sheet in our study is linearly stable with respect to tearing modes, and the plasmoid instability develops as a secondary instability as a sequel to the 3D nonlinear ballooning instability. 

The reconnection process in our simulations is not invariant in the cross-tail direction (that is, the $y$-direction). This leads to the general question of how to characterize 3D reconnection in a magnetotail that does not appear to contain a 2D X-line. Indeed, as is well known, a 2D X-line, which consists of a continuum of nulls, is structurally unstable and breaks up into a group of discrete nulls~\citep{lau90a}.

{\it Quasi-separatrix layer.} To address these questions in this work, we for the first time, apply the geometric concept of quasi-separatrix layer (QSL) to the analysis of the geometry of magnetic reconnection induced by ballooning instability in a generalized Harris sheet that represents the magneotail. The QSL has been adopted for the analysis of the reconnection structures involved in the solar corona processes for a long time (e.g.~\cite{titov99a,titov02a}). It has also been recently applied to the analysis of laboratory reconnection experiments~\cite{lawrence09a}. A QSL is a region with steep gradient in the field line connectivity. QSL's are constructed by mapping field lines across a specified volume. A surface, $S$, must first be defined to surround some volume of one's magnetic field. Divide $S$ into two subspaces, $S_0$ and $S_1$, where $S_0$ represents the surface on which field lines enter the volume, and $S_1$ represents those that leave. The initial footpoint is defined as $(u_0,v_0)$ in $S_0$. One then traces the field line from the initial footpoint through the enclosed volume until the field line leaves the volume through $S_1$ at the point $(u_1,v_1)$. The Jacobian transformation matrix and the norm of the mapping from $(u_0,v_0)$ to $(u_1,v_1)$ are defined as 
\begin{eqnarray}
{\cal J}&=&
\left(\begin{array}{cc}
\frac{\partial u_1}{\partial u_0} & \frac{\partial u_1}{\partial v_0} \\
\frac{\partial v_1}{\partial u_0} & \frac{\partial v_1}{\partial v_0}
\end{array}\right) \\
N&=&\sqrt{\left(\frac{\partial u_1}{\partial u_0}\right)^2+\left(\frac{\partial u_1}{\partial v_0}\right)^2+\left(\frac{\partial v_1}{\partial u_0}\right)^2+\left(\frac{\partial v_1}{\partial v_0}\right)^2}.
\label{N_eqn}
\end{eqnarray}
A QSL is the region where the gradient of this mapping is large compared to the average mapping, i.e. $N>>1$.

Mathematically, the squashing degree $Q$ is defined as $Q=N^2/|\Delta|$ where $\Delta$ is the determinant of the Jacobian matrix~\cite{titov02a,priest95a}. The variation of $Q$ among different field lines reflects the deformation of the magnetic flux tubes. A high squashing degree corresponds to a contraction in the cross-sectional area of a flux tube. The 2D projections of quasi-separatrix layers turn into separatrices in the limit the layer thickness goes to zero, or the corresponding squashing degree goes to infinity. The physical significance of QSL is that current sheets may form on these layers for reconnection.

{\it Bald patch.} Another closely related geometric concept is that of a bald patch (BP)~\cite{titov93a,titov99a}. Mathematically a BP is defined as the region of a magnetic field where $(\textbf{B}_\perp \cdot \nabla_\perp B_n)|_{IL}>0$. Here $\perp$ corresponds to the horizontal components within the mapping plane, the direction $n$ is taken normal to the mapping plane, and $IL$ denotes the inversion line where the BP condition is evaluated. The concept of BP has long been applied to the analyses of the magnetic field structure on solar photosphere~\cite{titov99a,titov02a}. From the mathematical definitions of BP's, QSL's, and squashing degrees, we can expect the formation of quasi-separatrix layers around BP's~\cite{demoulin96a}. As shown later, the boundary of a BP, as defined in $(\textbf{B}_\perp \cdot \nabla_\perp B_n)|_{IL}=0$, can be used to designate the location of QSLs formed during the 3D magnetic reconnection process in the generalized Harris sheet as well. 

In this work, we compute both the bald patches and the squashing degrees to identify the QSLs of the magnetic field configuration associated with the plasmoid formation, in an attempt to understand the global geometry of the magnetic field and the 3D nature of the magnetic reconnection process induced by ballooning instability.

\begin{figure}
\includegraphics[width=0.49\textwidth]{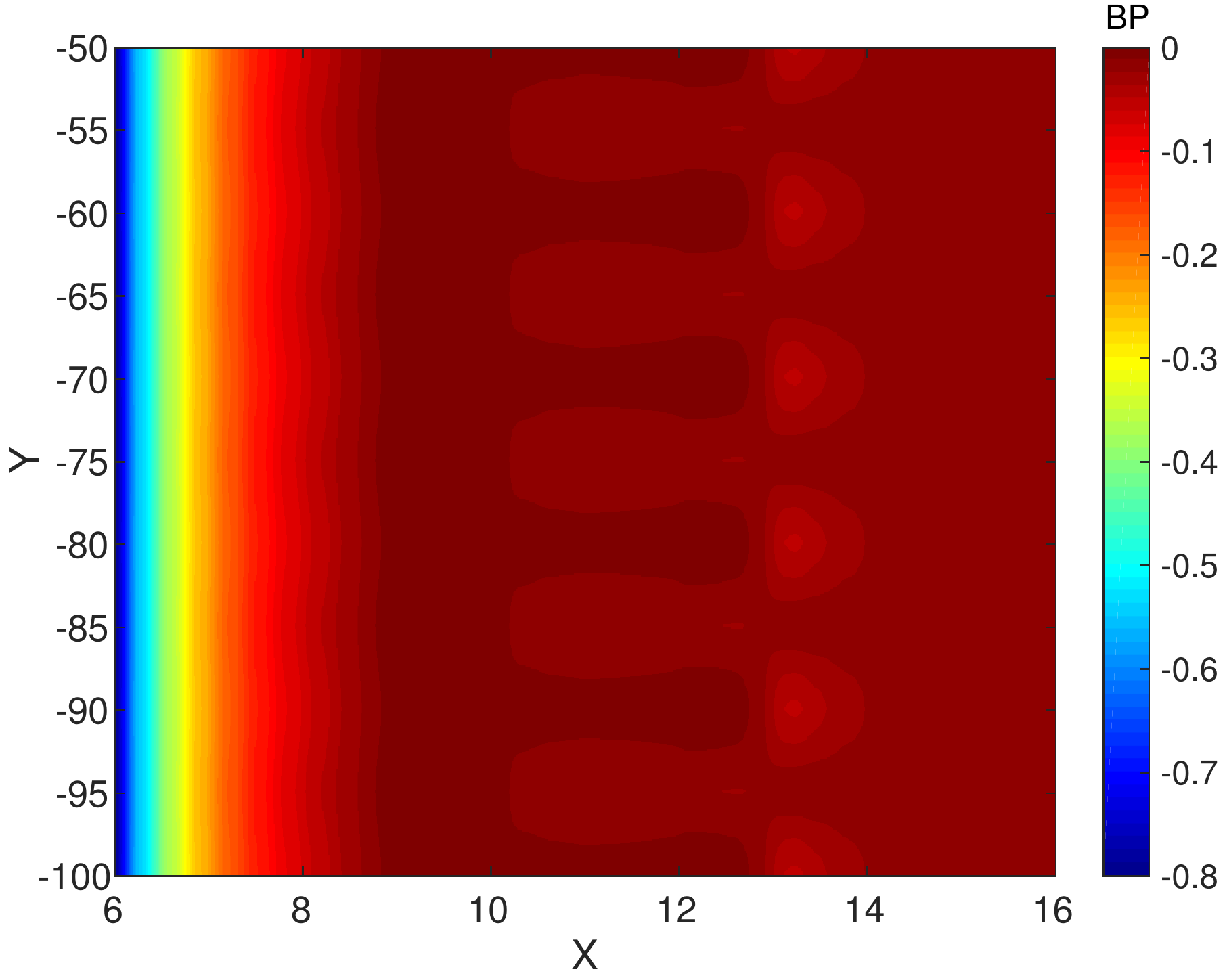}
\includegraphics[width=0.49\textwidth]{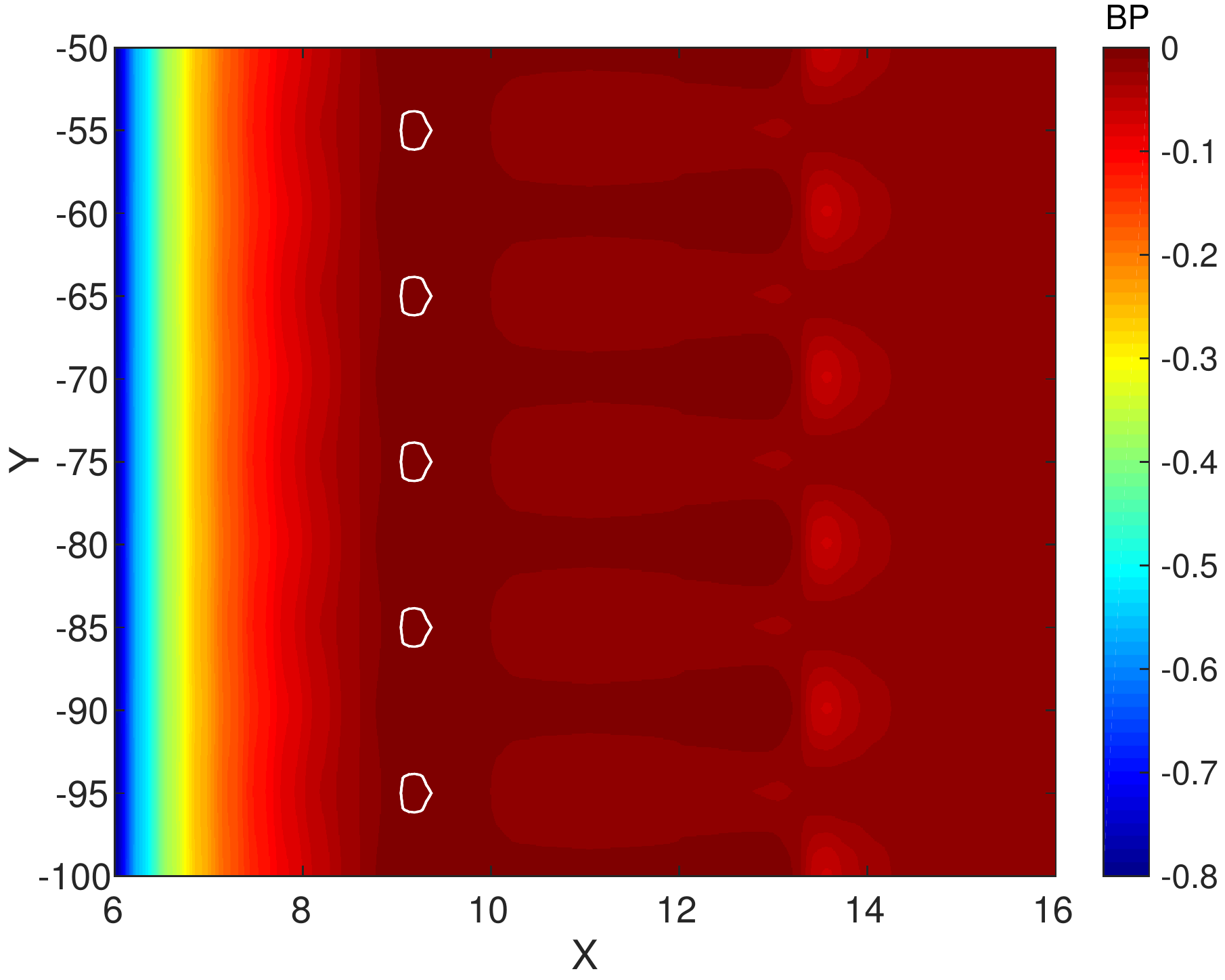}
\includegraphics[width=0.49\textwidth]{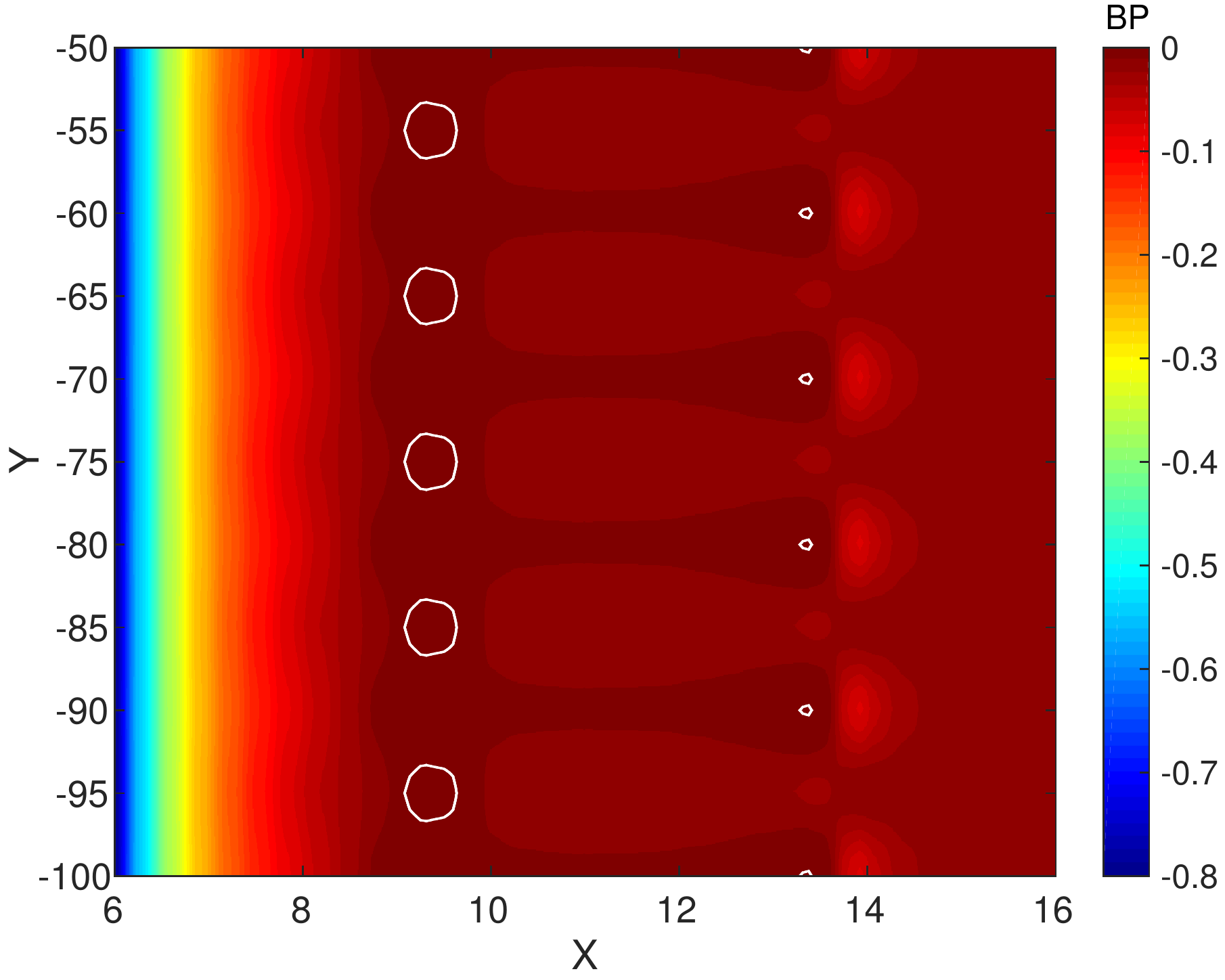}
\includegraphics[width=0.49\textwidth]{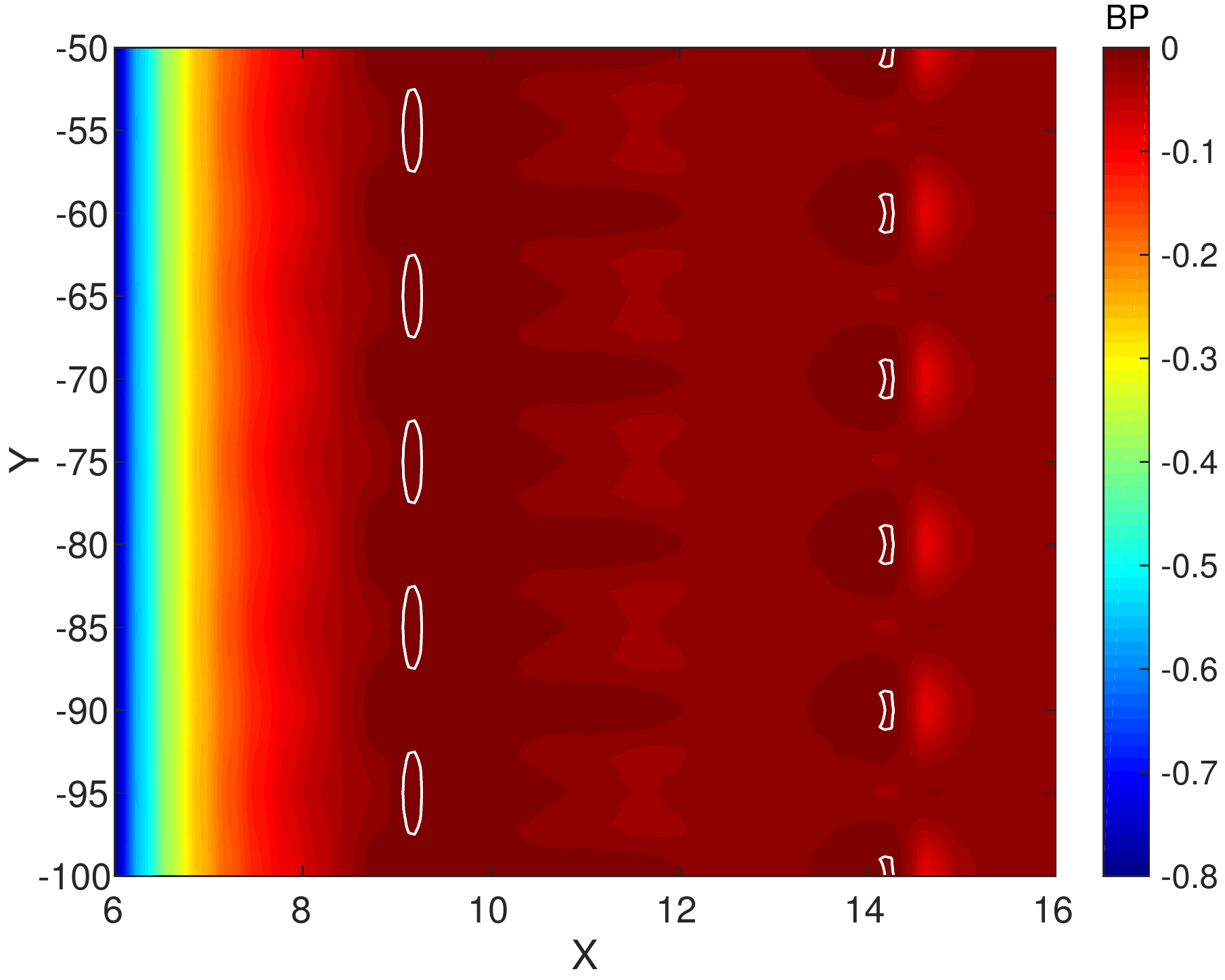}
\includegraphics[width=0.49\textwidth]{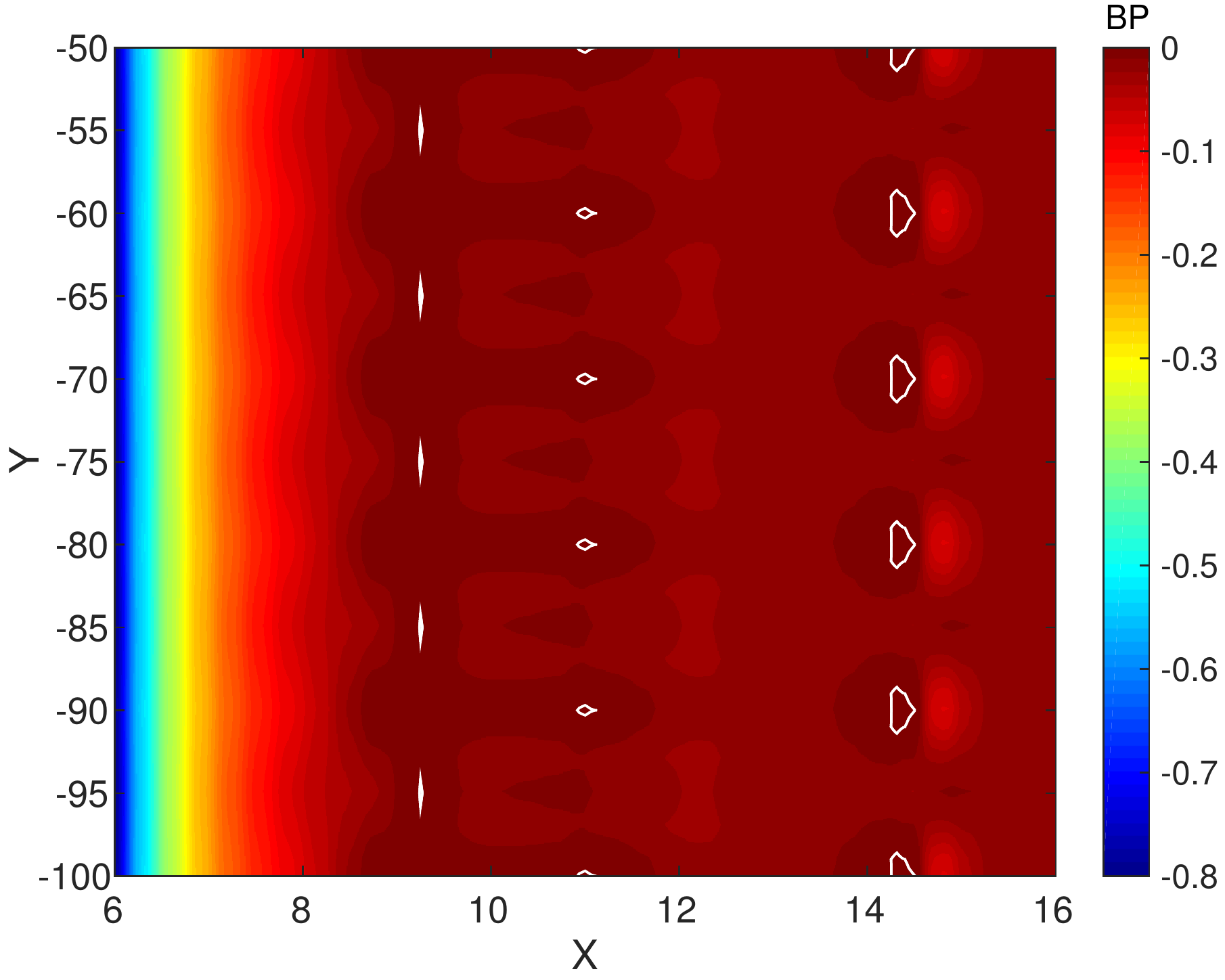}
\includegraphics[width=0.49\textwidth]{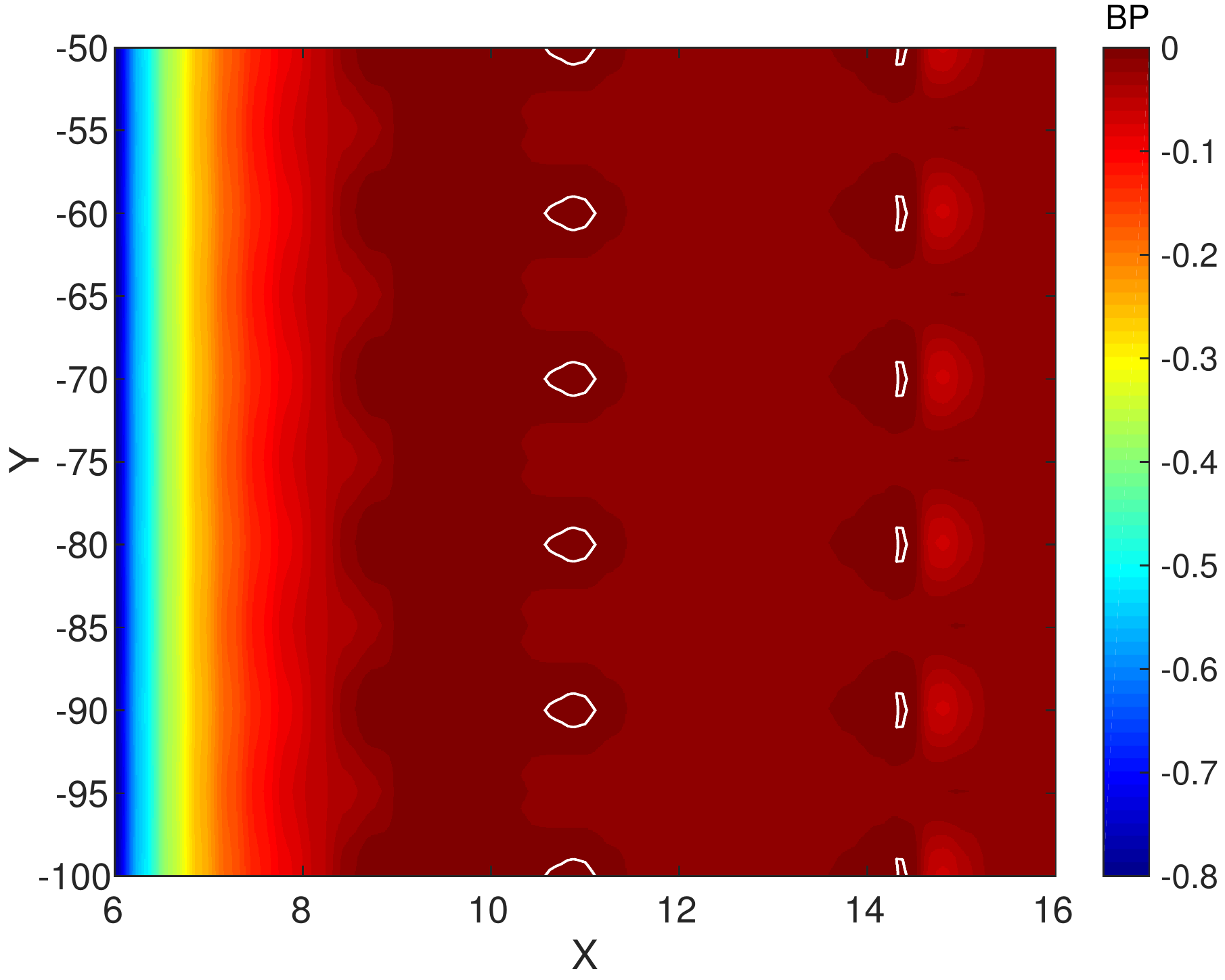}
\caption{Contours of $(\textbf{B}_\perp \cdot \nabla_\perp B_n)|_{IL}$ in $z=0$ plane at $t=170$ (upper left), $t=180$ (upper right), $t=190$ (middle left), $t=220$ (middle right), $t=240$ (lower left), $t=260$ (lower right). White circles denote the locations where $(\textbf{B}_\perp \cdot \nabla_\perp B_n)|_{IL}=0$. BP's are inside the white circles where $(\textbf{B}_\perp \cdot \nabla_\perp B_n)|_{IL}>0$.}
\label{fig:bp_t17_t26}
\end{figure}

We first briefly describe the development of bald patches in the inversion plane of the current sheet (i.e. $z=0$ plane) during the course of the ballooning instability evolution, as shown in Fig.~\ref{fig:bp_t17_t26}. The inversion plane would correspond to the equatorial plane in a model for the near-Earth plasma sheet. In the initial and early stage of ballooning instability evolution, bald patches are absent in the $z=0$ plane ($t=170$) (Fig.~\ref{fig:bp_t17_t26}, upper left). By the time $t=180$ the first set of bald patches shown as the region enclosed by their boundaries depicted as the white circles start to form periodically along the $y$ direction within the $z=0$ plane around the line of $x=9.5$ (Fig.~\ref{fig:bp_t17_t26}, upper right). 

As the ballooning instability continues to evolve, a second set of bald patches start to form in the equatorial plane near the radially extending fronts of ballooning fingers around $x\lesssim 13.5$ ($t=190$) (Fig.~\ref{fig:bp_t17_t26}, middle left). The circular shape of each of these BP's is smaller in radius. Their spatial distribution pattern is similar to the first set of BP circles, but their locations are shifted in $y$ direction from the first set by one half distance between two adjacent BP circles. After reaching their maximum sizes, the first set of BP circles begin to shrink into ellipses squeezed in the $x$ direction and eventually disappear ($t=220-260$) (Fig.~\ref{fig:bp_t17_t26}, middle right, lower left, and lower right). In addition, the locations of the BP circles also evolve, particularly for the second set. As the ballooning finger tips extend in the positive $x$ direction, the BP circles behind the each finger tip in the second set move along in the same direction.

Furthermore, as the first set of BP circles nearly disappear, a third set of BP circles start to emerge at $x=11$ between the first two sets around $t=240$ (Fig.~\ref{fig:bp_t17_t26}, lower left). This set of BP circles later become dominant in size after the first set disappear and the second set also shrinks in size. Different from the first set, the third set of BP circles have the same locations in $y$ as those in the second set.

\begin{figure}
\includegraphics[width=0.49\textwidth]{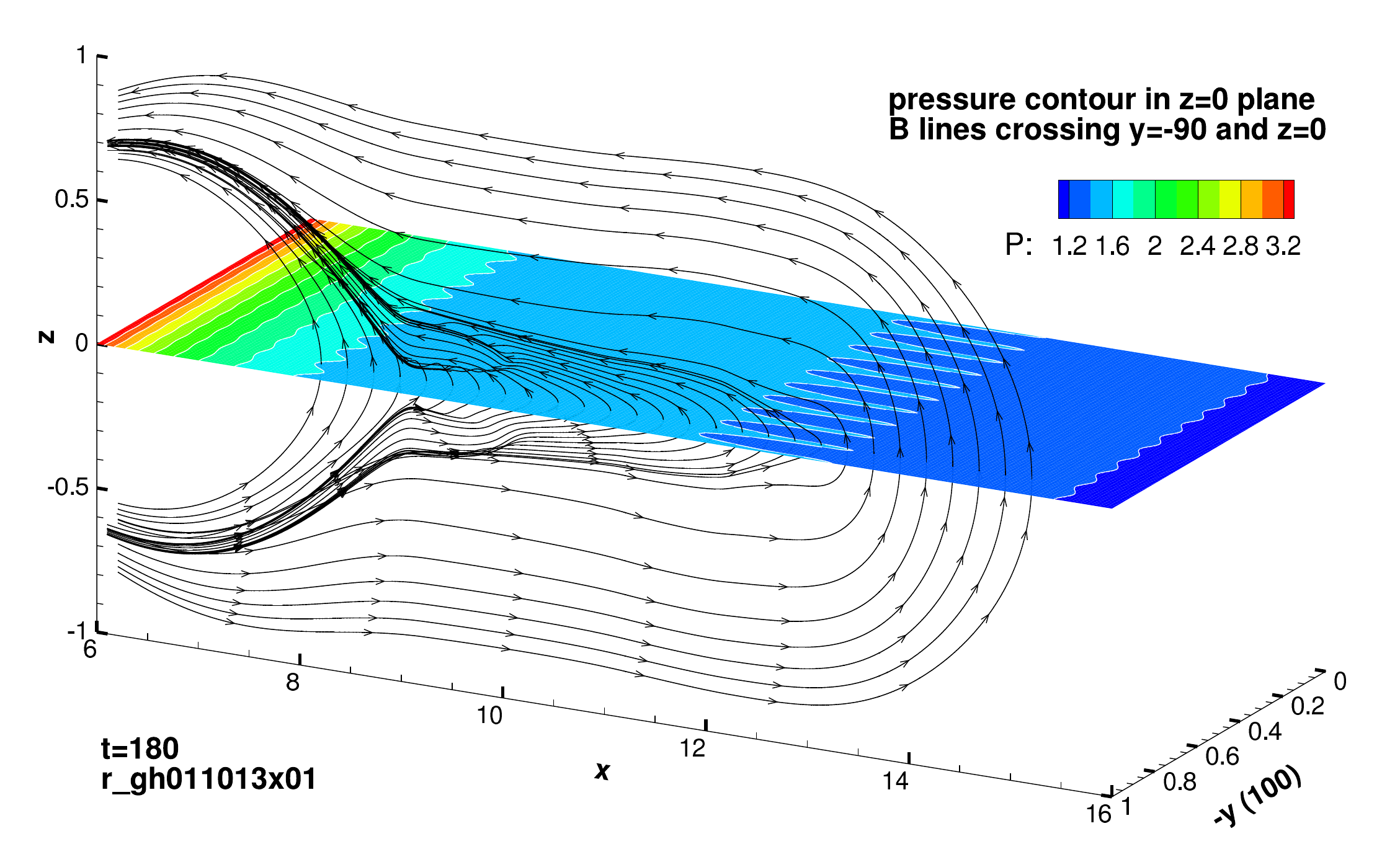}
\includegraphics[width=0.49\textwidth]{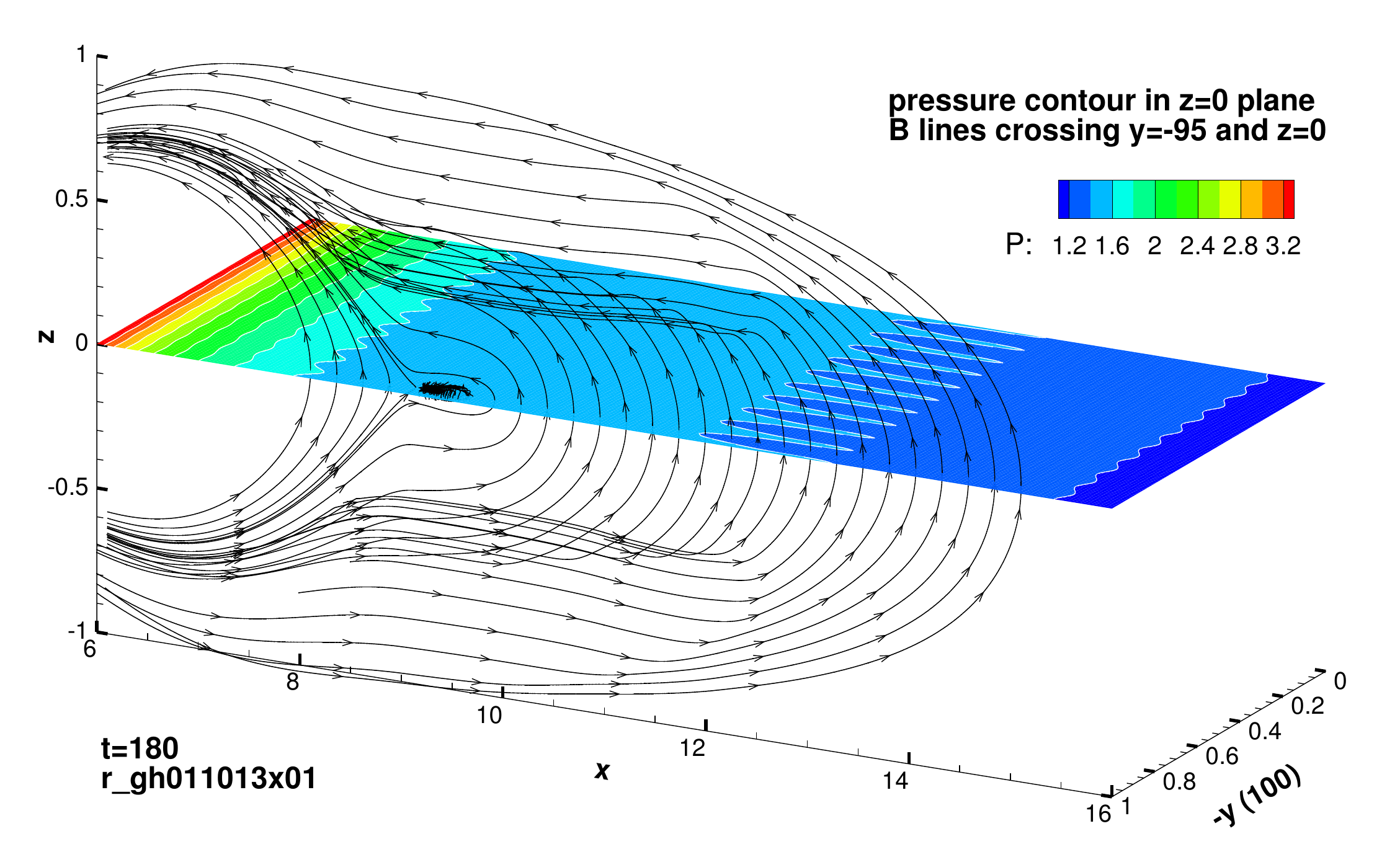}
\includegraphics[width=0.49\textwidth]{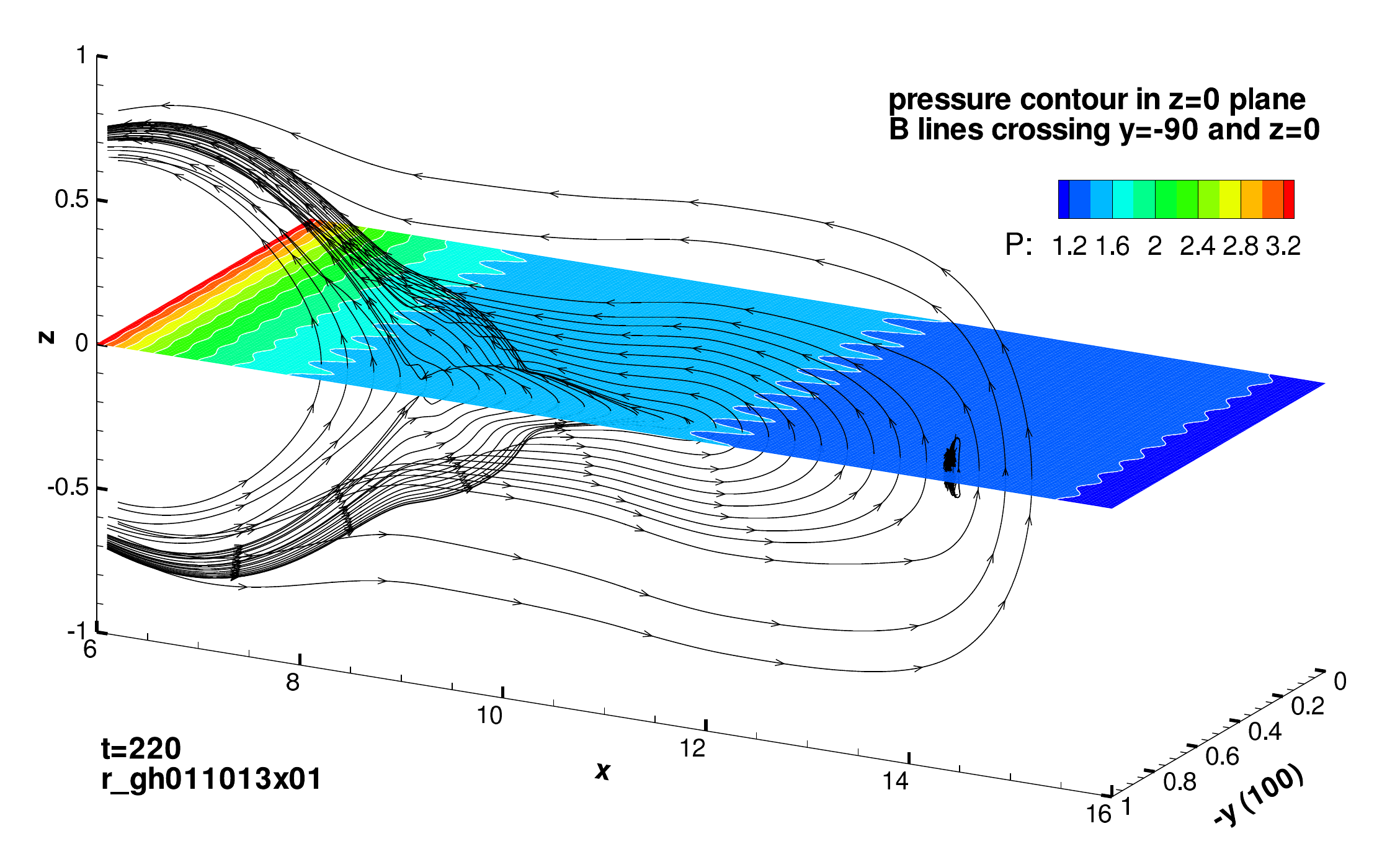}
\includegraphics[width=0.49\textwidth]{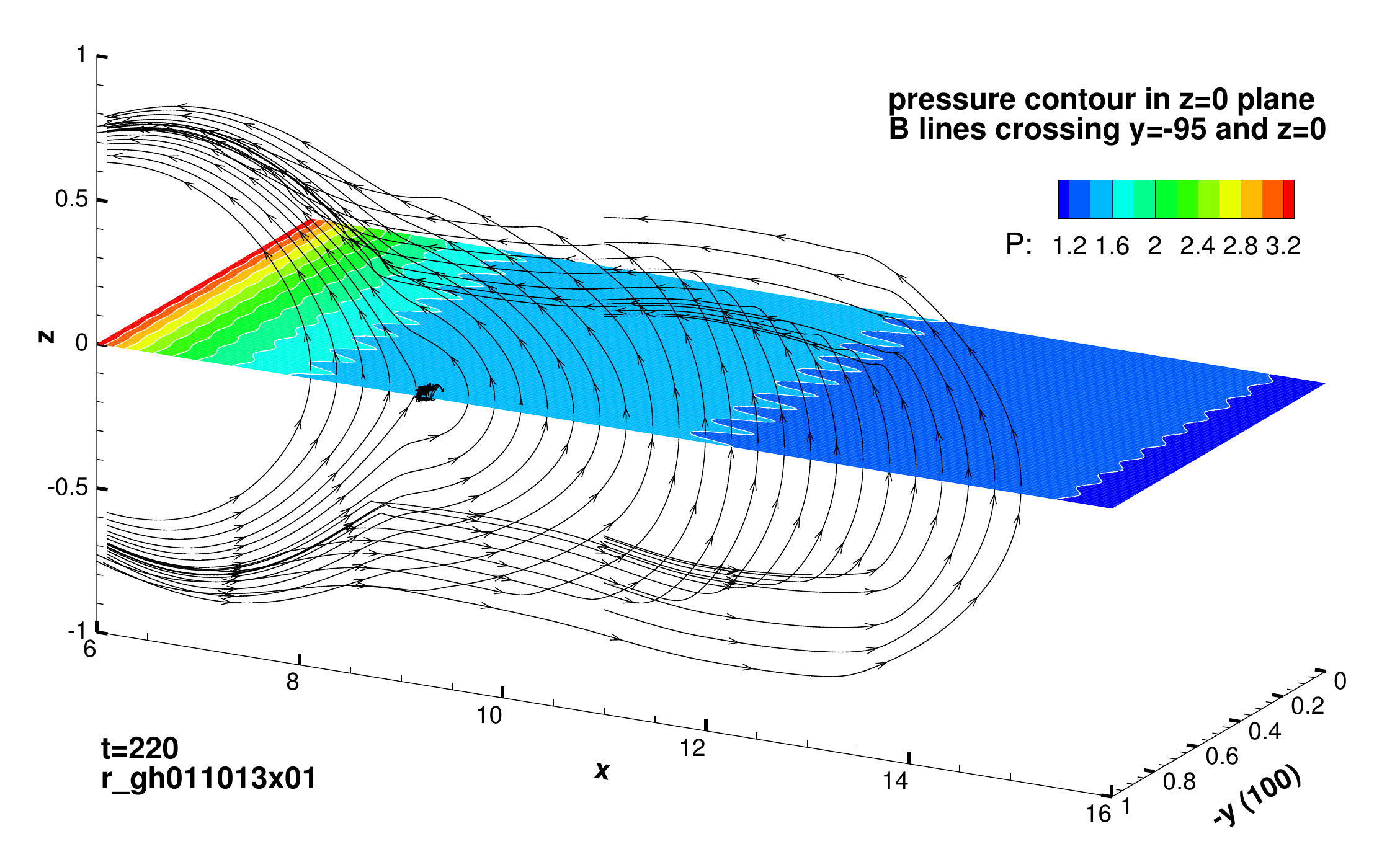}
\includegraphics[width=0.49\textwidth]{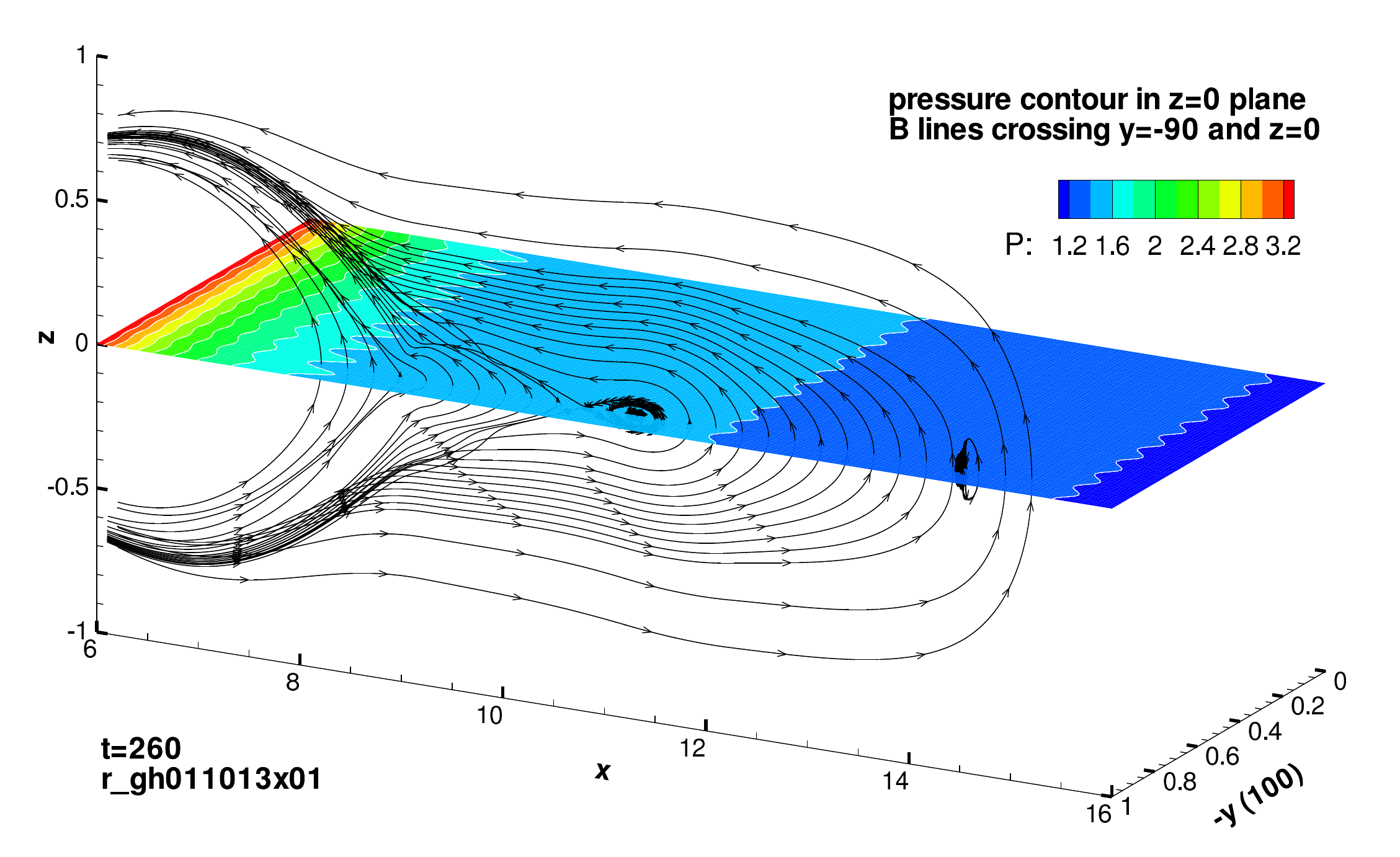}
\includegraphics[width=0.49\textwidth]{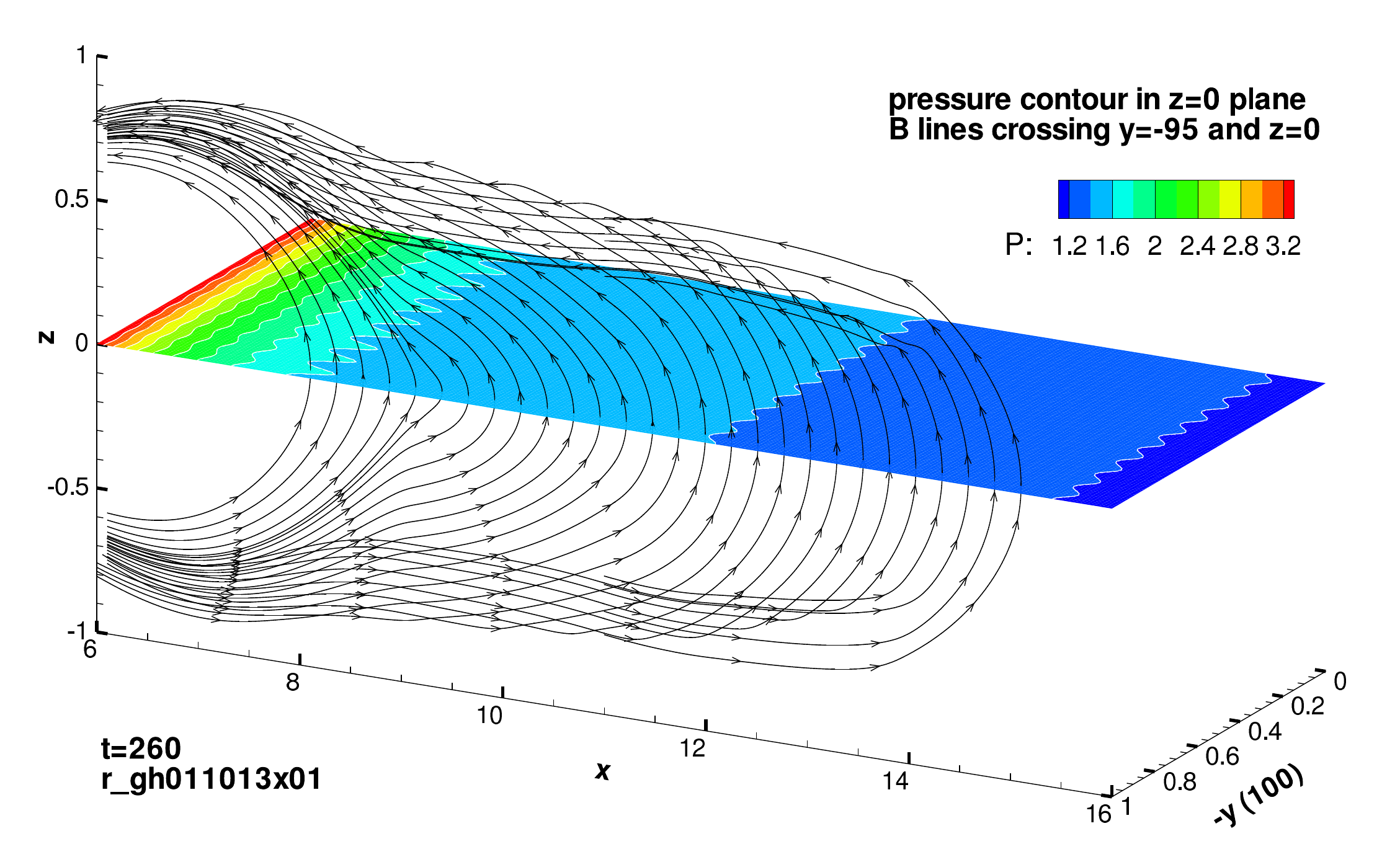}
\caption{Magnetic field lines crossing the line $y=-90, z=0$ (left) and the line $y=-95, z=0$ (right) at $t=180$ (upper), $t=220$ (middle), and $t=260$ (lower).}
\label{fig:bline_y90_t18_t19}
\end{figure}

The spatial distribution of the QSL's indicated by these BP circles reveals the global 3D geometry of the associated plasmoid formation and reconnection processes. In particular, the locations of the QSL's are exactly where the plasmoids are found to develop. Take the time slices of $t=180, 220, 260$ for example, which correspond to the time moments of the panels in the right column of Fig.~\ref{fig:bp_t17_t26}. At each of the three moments, we trace the magnetic field lines from the line $y=-90, z=0$ and the line $y=-95, z=0$, and show the field line structure in the left and the right columns of Fig.~\ref{fig:bline_y90_t18_t19}, respectively.

For field lines that cross the equatorial plane at points away from the QSL's, the field line structure and connectivity remain topologically the same as in the beginning. One such case is shown in Fig.~\ref{fig:bline_y90_t18_t19} (upper left) where all the field lines cross the $z=0$ plane through the line $y=-90,z=0$ at $t=180$. After the QSL's appear near the ballooning finger tip around $x=13.5,y=-90,-80,\cdots$ in the $z=0$ plane at $t=220$, a small plasmoid is spotted to form around $x=13.5$ on those same field lines across the $y=-90, z=0$ line (Fig.~\ref{fig:bline_y90_t18_t19}, middle left). At a later time ($t=260$), whereas that plasmoid has moved slightly toward the positive $x$-direction, a larger plasmoid appears on the same line of $y=-90$ in the $z=0$ plane, but at a location around $x=10.5$ much behind that of the smaller plasmoid (Fig.~\ref{fig:bline_y90_t18_t19}, lower left). The BP plot at the same time shows the emergence of a set of larger circle-shaped QSL's around the line $x=10.5, z=0$ along the $y$ direction (Fig.~\ref{fig:bp_t17_t26}, lower right). Similarly, for those field lines crossing the line $y=-95,z=0$, which is half wavelength of the dominant linear ballooning instability in the $y$-direction away from the field lines in the previous case in the $z=0$ plane, the appearance of plasmoids in Fig.~\ref{fig:bline_y90_t18_t19} (right column) conforms exactly with the emergence of QSL's as indicated by the BP contours in Fig.~\ref{fig:bp_t17_t26} (right column) in both time and location. 

\begin{figure}
\includegraphics[width=0.49\textwidth]{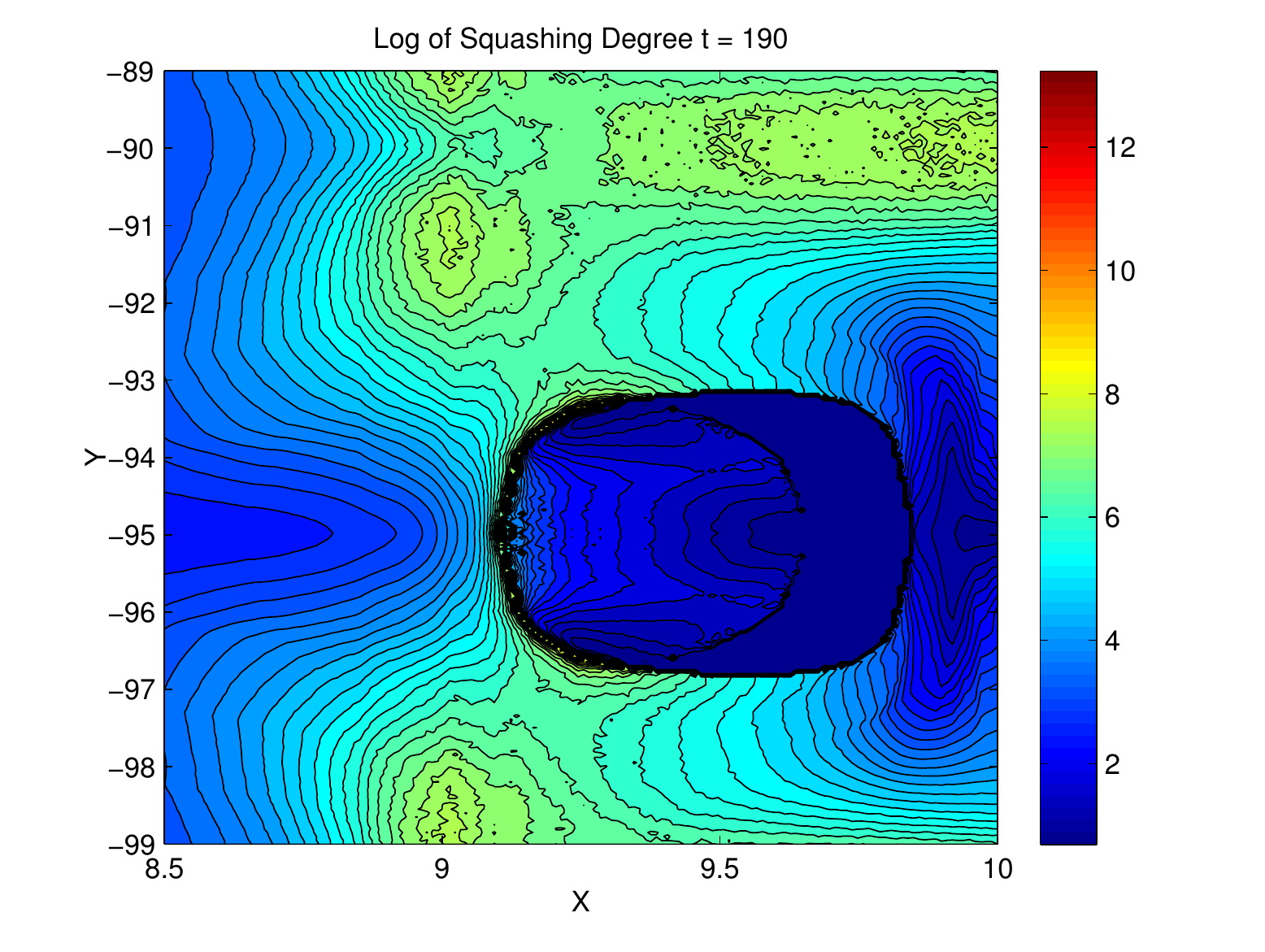}
\includegraphics[width=0.49\textwidth]{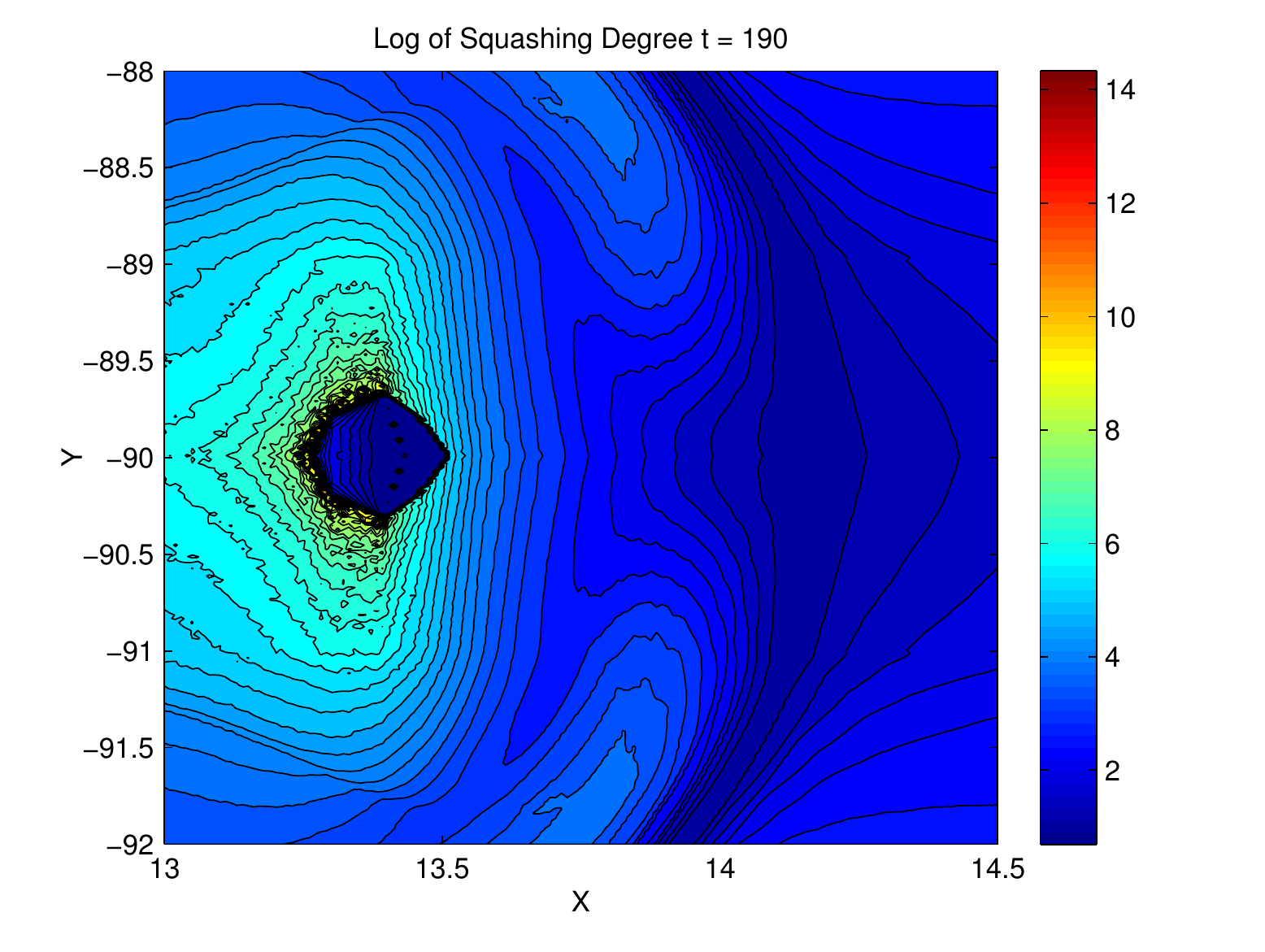}
\caption{Contours of the logarithm of squashing degree in the $z=0$ plane around $x=9.5, y=-95$ (left) and $x=13.4, y=-90$ (right) at $t=190$.} 
\label{fig:sq_t19}
\end{figure}

Our calculations find that the QSL's identified from the BP contours agree with those from the contours of squashing degrees. For example, the QSL's represented by the dark-lined circles in Fig.~\ref{fig:sq_t19} which are centered around $x=9.5, y=-95$ and $x=13.4, y=-90$. They have essentially the same shapes and locations as the QSL's represented by the white circles in Fig.~\ref{fig:bp_t17_t26} for the same time moment ($t=19$). The size and shape of the QSL's correspond to those of the plasmoids in the 2D $x-z$ plane as projected to from the flux ropes in 3D.

The periodic distribution of QSL's in the $y$-direction indicates that the underlying reconnection process cannot take place in a 2D domain where $y$ is a direction of continuous symmetry. The intrinsic 3D nature of the reconnecting field line geometry originates from the intrinsic 3D dynamics of the reconnection process. As shown previously, the formation of plasmoids in the generalized Harris sheet considered here is induced by the nonlinear development of ballooning instability~\cite{zhu13d,zhu14a}. Because the ballooning instability can only grow into nonlinear stage with a finite wavenumber in $y$-direction (i.e. $k_y\neq 0$), the dynamics of plasmoid formation and reconnection process induced by the nonlinear ballooning instability is inherently 3D in nature and cannot take place in any 2D configuration. In other words, there is no equivalent 2D reconnection process that the 3D reconnection described in this paper can reduce to. It may be argued that our considerations have been based on the resistive MHD model, and are hence inapplicable to the collisionless magnetotail. We remark that previous studies have demonstrated convincingly the development of the ballooning instability in a generalized Harris sheet using both two-fluid and kinetic models~\citep{bhattacharjee98a,cheng98a,zhu03a,pritchett13a}. Hence the occurrence of the ballooning instability is not in doubt. Furthermore, our general considerations pertaining to the global topology of magnetic field lines are not very sensitive to the precise mechanisms that break field lines. In other words, our present model captures the qualitative features of global evolution reasonably accurately.

In summary, a magnetic reconnection process in the generalized Harris sheet has been revealed to be intrinsically 3-dimensional both geometrically and dynamically despite the spatial invariance of the initial current sheet in the dawn-dusk (or $y$) direction. The intrinsic geometrical 3-dimensionality of the reconnection process is a direct consequence of the intrinsic 3D nature of its MHD driver, i.e., the ballooning instability. Due to the ubiquitous presence of 3D MHD instabilities and their associated reconnection processes in space, laboratory, and astrophysical plasmas, our study is an object lesson on the need to go beyond the commonly adopted paradigm of 2D magnetic reconnection. In particular, our finding provides a new perspective to the long-standing controversy over the substorm onset problem, since it naturally unifies the traditional reconnection based and ballooning instability based scenarios through the 3D reconnection process.

Our work also connects to fundamental developments in the theory of 3D reconnection processes in solar as well as laboratory plasmas. For example, previous simulations demonstrate the nonlinear growth of double tearing mode driven by kinetic ballooning turbulence in a tokamak~\citep{ishizawa07a}. There the 2D tokamak equilibrium configuration features intrinsic magnetic shear, which are linearly unstable to tearing modes on rational surfaces. This contrasts with the magnetotail configuration represented by the generalized Harris sheet considered in our work, which has no magnetic shear or rational surface, and is linearly stable to tearing modes for the weakly collisional and collisionless regimes ($S\ge 10^4$). Despite these differences, our work offers a common perspective for 3D reconnection underlying seemingly different phenomena.

The rate of reconnection and energy conversion in 3D are open problems in reconnection physics that merit a separate investigation, and are well beyond the scope of this Brief Communication. We plan on studying these issues in our future work.

\begin{acknowledgments}
This research was supported by Natural Science Foundation of China Grant No. 41474143, U.S. NSF Grant No. AGS-0902360, the 100 Talent Program of Chinese Academy of Sciences, U.S. DOE Grant Nos. DE-FG02-86ER53218 and DE-FC02-08ER54975. The computational work used the XSEDE resources (U.S. NSF Grant No. ACI-1053575) provided by TACC under Grant No. TG-ATM070010, and the resources of NERSC, which is supported by U.S. DOE under Contract No. DE-AC02-05CH11231. A. B. would like to acknowledge U.S. NSF Grant Nos. AGS-1338944 and AGS-1460169.
\end{acknowledgments}

\newpage
\clearpage

\clearpage

\clearpage

\begin{thebibliography}{25}
\expandafter\ifx\csname natexlab\endcsname\relax\def\natexlab#1{#1}\fi
\expandafter\ifx\csname bibnamefont\endcsname\relax
  \def\bibnamefont#1{#1}\fi
\expandafter\ifx\csname bibfnamefont\endcsname\relax
  \def\bibfnamefont#1{#1}\fi
\expandafter\ifx\csname citenamefont\endcsname\relax
  \def\citenamefont#1{#1}\fi
\expandafter\ifx\csname url\endcsname\relax
  \def\url#1{\texttt{#1}}\fi
\expandafter\ifx\csname urlprefix\endcsname\relax\def\urlprefix{URL }\fi
\providecommand{\bibinfo}[2]{#2}
\providecommand{\eprint}[2][]{\url{#2}}

\bibitem[{\citenamefont{Sweet}(1958{\natexlab{a}})}]{sweet58a}
\bibinfo{author}{\bibfnamefont{P.~A.} \bibnamefont{Sweet}}, in
  \emph{\bibinfo{booktitle}{Electromagnetic Phenomena in Cosmical Physics,
  Proceedings from International Astronomical Union Symposium}}
  (\bibinfo{publisher}{Cambridge University Press}, \bibinfo{address}{London},
  \bibinfo{year}{1958}{\natexlab{a}}), vol.~\bibinfo{volume}{6}, p.
  \bibinfo{pages}{123}, \bibinfo{note}{edited by Bo Lehnert}.

\bibitem[{\citenamefont{Sweet}(1958{\natexlab{b}})}]{sweet58b}
\bibinfo{author}{\bibfnamefont{P.~A.} \bibnamefont{Sweet}},
  \bibinfo{journal}{Nuovo Cimento Suppl.} \textbf{\bibinfo{volume}{8}},
  \bibinfo{pages}{Ser. X, 188} (\bibinfo{year}{1958}{\natexlab{b}}).

\bibitem[{\citenamefont{Parker}(1963)}]{parker63a}
\bibinfo{author}{\bibfnamefont{E.~N.} \bibnamefont{Parker}},
  \bibinfo{journal}{Astrophys. J. Suppl. Ser.} \textbf{\bibinfo{volume}{8}},
  \bibinfo{pages}{177} (\bibinfo{year}{1963}).

\bibitem[{\citenamefont{Petschek}(1964)}]{petschek64a}
\bibinfo{author}{\bibfnamefont{H.~E.} \bibnamefont{Petschek}}, in
  \emph{\bibinfo{booktitle}{AAS/NASA Symposium on the Physics of Solar Flares,
  ed. W. N. Hess}} (\bibinfo{publisher}{NASA}, \bibinfo{address}{Washington,
  DC}, \bibinfo{year}{1964}), pp. \bibinfo{pages}{425--37}.

\bibitem[{\citenamefont{Pontin}(2011)}]{pontin11a}
\bibinfo{author}{\bibfnamefont{D.~I.} \bibnamefont{Pontin}},
  \bibinfo{journal}{{\it Adv. Space Res.}} \textbf{\bibinfo{volume}{47}},
  \bibinfo{pages}{1508} (\bibinfo{year}{2011}).

\bibitem[{\citenamefont{Angelopoulos et~al.}(2008)\citenamefont{Angelopoulos,
  McFadden, Larson, Carlson, Mende, Frey, Phan, Sibeck, Glassmeier, Auster
  et~al.}}]{angelopoulos08c}
\bibinfo{author}{\bibfnamefont{V.}~\bibnamefont{Angelopoulos}},
  \bibinfo{author}{\bibfnamefont{J.}~\bibnamefont{McFadden}},
  \bibinfo{author}{\bibfnamefont{D.}~\bibnamefont{Larson}},
  \bibinfo{author}{\bibfnamefont{C.}~\bibnamefont{Carlson}},
  \bibinfo{author}{\bibfnamefont{S.}~\bibnamefont{Mende}},
  \bibinfo{author}{\bibfnamefont{H.}~\bibnamefont{Frey}},
  \bibinfo{author}{\bibfnamefont{T.}~\bibnamefont{Phan}},
  \bibinfo{author}{\bibfnamefont{D.}~\bibnamefont{Sibeck}},
  \bibinfo{author}{\bibfnamefont{K.-H.} \bibnamefont{Glassmeier}},
  \bibinfo{author}{\bibfnamefont{U.}~\bibnamefont{Auster}},
  \bibnamefont{et~al.}, \bibinfo{journal}{Science}
  \textbf{\bibinfo{volume}{321}}, \bibinfo{pages}{931} (\bibinfo{year}{2008}).

\bibitem[{\citenamefont{Panov et~al.}(2012)\citenamefont{Panov, Sergeev,
  Pritchett, Coroniti, Nakamura, Baumjohann, Angelopoulos, Auster, and
  McFadden}}]{panov12a}
\bibinfo{author}{\bibfnamefont{E.~V.} \bibnamefont{Panov}},
  \bibinfo{author}{\bibfnamefont{V.~A.} \bibnamefont{Sergeev}},
  \bibinfo{author}{\bibfnamefont{P.~L.} \bibnamefont{Pritchett}},
  \bibinfo{author}{\bibfnamefont{F.~V.} \bibnamefont{Coroniti}},
  \bibinfo{author}{\bibfnamefont{R.}~\bibnamefont{Nakamura}},
  \bibinfo{author}{\bibfnamefont{W.}~\bibnamefont{Baumjohann}},
  \bibinfo{author}{\bibfnamefont{V.}~\bibnamefont{Angelopoulos}},
  \bibinfo{author}{\bibfnamefont{H.~U.} \bibnamefont{Auster}},
  \bibnamefont{and} \bibinfo{author}{\bibfnamefont{J.~P.}
  \bibnamefont{McFadden}}, \bibinfo{journal}{\grl}
  \textbf{\bibinfo{volume}{39}}, \bibinfo{pages}{L08110}
  (\bibinfo{year}{2012}).

\bibitem[{\citenamefont{Sovinec et~al.}(2004)\citenamefont{Sovinec, Glasser,
  Barnes, Gianakon, Nebel, Kruger, Schnack, Plimpton, Tarditi, Chu
  et~al.}}]{sovinec04a}
\bibinfo{author}{\bibfnamefont{C.}~\bibnamefont{Sovinec}},
  \bibinfo{author}{\bibfnamefont{A.}~\bibnamefont{Glasser}},
  \bibinfo{author}{\bibfnamefont{D.}~\bibnamefont{Barnes}},
  \bibinfo{author}{\bibfnamefont{T.}~\bibnamefont{Gianakon}},
  \bibinfo{author}{\bibfnamefont{R.}~\bibnamefont{Nebel}},
  \bibinfo{author}{\bibfnamefont{S.}~\bibnamefont{Kruger}},
  \bibinfo{author}{\bibfnamefont{D.}~\bibnamefont{Schnack}},
  \bibinfo{author}{\bibfnamefont{S.}~\bibnamefont{Plimpton}},
  \bibinfo{author}{\bibfnamefont{A.}~\bibnamefont{Tarditi}},
  \bibinfo{author}{\bibfnamefont{M.}~\bibnamefont{Chu}}, \bibnamefont{et~al.},
  \bibinfo{journal}{\jcp} \textbf{\bibinfo{volume}{195}}, \bibinfo{pages}{355}
  (\bibinfo{year}{2004}).

\bibitem[{\citenamefont{Zhu and Raeder}(2013)}]{zhu13d}
\bibinfo{author}{\bibfnamefont{P.}~\bibnamefont{Zhu}} \bibnamefont{and}
  \bibinfo{author}{\bibfnamefont{J.}~\bibnamefont{Raeder}},
  \bibinfo{journal}{\prl} \textbf{\bibinfo{volume}{110}},
  \bibinfo{pages}{235005} (\bibinfo{year}{2013}).

\bibitem[{\citenamefont{Zhu and Raeder}(2014)}]{zhu14a}
\bibinfo{author}{\bibfnamefont{P.}~\bibnamefont{Zhu}} \bibnamefont{and}
  \bibinfo{author}{\bibfnamefont{J.}~\bibnamefont{Raeder}},
  \bibinfo{journal}{{\it J. Geophys. Res. Space Physics}}
  \textbf{\bibinfo{volume}{119}}, \bibinfo{pages}{131} (\bibinfo{year}{2014}).

\bibitem[{\citenamefont{Birn and {Hones, Jr.}}(1981)}]{birn81a}
\bibinfo{author}{\bibfnamefont{J.}~\bibnamefont{Birn}} \bibnamefont{and}
  \bibinfo{author}{\bibfnamefont{E.~W.} \bibnamefont{{Hones, Jr.}}},
  \bibinfo{journal}{\jgr} \textbf{\bibinfo{volume}{86}},
  \bibinfo{pages}{6802–} (\bibinfo{year}{1981}).

\bibitem[{\citenamefont{Hesse and Birn}(1991)}]{hesse91a}
\bibinfo{author}{\bibfnamefont{M.}~\bibnamefont{Hesse}} \bibnamefont{and}
  \bibinfo{author}{\bibfnamefont{J.}~\bibnamefont{Birn}},
  \bibinfo{journal}{\jgr} \textbf{\bibinfo{volume}{96}},
  \bibinfo{pages}{5683–5696} (\bibinfo{year}{1991}).

\bibitem[{\citenamefont{Schindler}(2007)}]{schindler07a}
\bibinfo{author}{\bibfnamefont{K.}~\bibnamefont{Schindler}},
  \emph{\bibinfo{title}{Physics of Space Plasma Activity}}
  (\bibinfo{publisher}{Cambridge University Press},
  \bibinfo{address}{Cambridge, UK}, \bibinfo{year}{2007}).

\bibitem[{\citenamefont{Lau and Finn}(1990)}]{lau90a}
\bibinfo{author}{\bibfnamefont{Y.~T.} \bibnamefont{Lau}} \bibnamefont{and}
  \bibinfo{author}{\bibfnamefont{J.~M.} \bibnamefont{Finn}},
  \bibinfo{journal}{\apj} \textbf{\bibinfo{volume}{350}}, \bibinfo{pages}{672}
  (\bibinfo{year}{1990}).

\bibitem[{\citenamefont{Titov and D\'{e}moulin}(1999)}]{titov99a}
\bibinfo{author}{\bibfnamefont{V.~S.} \bibnamefont{Titov}} \bibnamefont{and}
  \bibinfo{author}{\bibfnamefont{P.}~\bibnamefont{D\'{e}moulin}},
  \bibinfo{journal}{{\it Astron. Astrophys.}} \textbf{\bibinfo{volume}{351}},
  \bibinfo{pages}{707} (\bibinfo{year}{1999}).

\bibitem[{\citenamefont{Titov and Hornig}(2002)}]{titov02a}
\bibinfo{author}{\bibfnamefont{V.~S.} \bibnamefont{Titov}} \bibnamefont{and}
  \bibinfo{author}{\bibfnamefont{G.}~\bibnamefont{Hornig}},
  \bibinfo{journal}{\jgr} \textbf{\bibinfo{volume}{107}}, \bibinfo{pages}{1164}
  (\bibinfo{year}{2002}).

\bibitem[{\citenamefont{Lawrence and Gekelman}(2009)}]{lawrence09a}
\bibinfo{author}{\bibfnamefont{E.~E.} \bibnamefont{Lawrence}} \bibnamefont{and}
  \bibinfo{author}{\bibfnamefont{W.}~\bibnamefont{Gekelman}},
  \bibinfo{journal}{\prl} \textbf{\bibinfo{volume}{103}},
  \bibinfo{pages}{105002} (\bibinfo{year}{2009}).

\bibitem[{\citenamefont{Priest and Demoulin}(1995)}]{priest95a}
\bibinfo{author}{\bibfnamefont{E.~R.} \bibnamefont{Priest}} \bibnamefont{and}
  \bibinfo{author}{\bibfnamefont{P.}~\bibnamefont{Demoulin}},
  \bibinfo{journal}{\jgr} \textbf{\bibinfo{volume}{100}},
  \bibinfo{pages}{23443} (\bibinfo{year}{1995}).

\bibitem[{\citenamefont{Titov et~al.}(1993)\citenamefont{Titov, Priest, and
  Demoulin}}]{titov93a}
\bibinfo{author}{\bibfnamefont{V.~S.} \bibnamefont{Titov}},
  \bibinfo{author}{\bibfnamefont{E.~R.} \bibnamefont{Priest}},
  \bibnamefont{and} \bibinfo{author}{\bibfnamefont{P.}~\bibnamefont{Demoulin}},
  \bibinfo{journal}{{\it Astron. Astrophys.}} \textbf{\bibinfo{volume}{276}},
  \bibinfo{pages}{564} (\bibinfo{year}{1993}).

\bibitem[{\citenamefont{Demoulin et~al.}(1996)\citenamefont{Demoulin, Henoux,
  Priest, and Mandrini}}]{demoulin96a}
\bibinfo{author}{\bibfnamefont{P.}~\bibnamefont{Demoulin}},
  \bibinfo{author}{\bibfnamefont{J.~C.} \bibnamefont{Henoux}},
  \bibinfo{author}{\bibfnamefont{E.~R.} \bibnamefont{Priest}},
  \bibnamefont{and} \bibinfo{author}{\bibfnamefont{C.~H.}
  \bibnamefont{Mandrini}}, \bibinfo{journal}{{\it Astron. and Astrophys.}}
  \textbf{\bibinfo{volume}{308}}, \bibinfo{pages}{643} (\bibinfo{year}{1996}).

\bibitem[{\citenamefont{Bhattacharjee et~al.}(1998)\citenamefont{Bhattacharjee,
  Ma, and Wang}}]{bhattacharjee98a}
\bibinfo{author}{\bibfnamefont{A.}~\bibnamefont{Bhattacharjee}},
  \bibinfo{author}{\bibfnamefont{Z.~W.} \bibnamefont{Ma}}, \bibnamefont{and}
  \bibinfo{author}{\bibfnamefont{X.}~\bibnamefont{Wang}},
  \bibinfo{journal}{\grl} \textbf{\bibinfo{volume}{25}}, \bibinfo{pages}{861}
  (\bibinfo{year}{1998}).

\bibitem[{\citenamefont{Cheng and Lui}(1998)}]{cheng98a}
\bibinfo{author}{\bibfnamefont{C.~Z.} \bibnamefont{Cheng}} \bibnamefont{and}
  \bibinfo{author}{\bibfnamefont{A.~T.~Y.} \bibnamefont{Lui}},
  \bibinfo{journal}{\grl} \textbf{\bibinfo{volume}{25}}, \bibinfo{pages}{4091}
  (\bibinfo{year}{1998}).

\bibitem[{\citenamefont{Zhu et~al.}(2003)\citenamefont{Zhu, Bhattacharjee, and
  Ma}}]{zhu03a}
\bibinfo{author}{\bibfnamefont{P.}~\bibnamefont{Zhu}},
  \bibinfo{author}{\bibfnamefont{A.}~\bibnamefont{Bhattacharjee}},
  \bibnamefont{and} \bibinfo{author}{\bibfnamefont{Z.~W.} \bibnamefont{Ma}},
  \bibinfo{journal}{\pop} \textbf{\bibinfo{volume}{10}}, \bibinfo{pages}{249}
  (\bibinfo{year}{2003}).

\bibitem[{\citenamefont{Pritchett and Coroniti}(2013)}]{pritchett13a}
\bibinfo{author}{\bibfnamefont{P.~L.} \bibnamefont{Pritchett}}
  \bibnamefont{and} \bibinfo{author}{\bibfnamefont{F.~V.}
  \bibnamefont{Coroniti}}, \bibinfo{journal}{\jgr}
  \textbf{\bibinfo{volume}{118}}, \bibinfo{pages}{146} (\bibinfo{year}{2013}).

\bibitem[{\citenamefont{Ishizawa and Nakajima}(2007)}]{ishizawa07a}
\bibinfo{author}{\bibfnamefont{A.}~\bibnamefont{Ishizawa}} \bibnamefont{and}
  \bibinfo{author}{\bibfnamefont{N.}~\bibnamefont{Nakajima}},
  \bibinfo{journal}{\pop} \textbf{\bibinfo{volume}{14}},
  \bibinfo{pages}{040702} (\bibinfo{year}{2007}).

\end{thebibliography}
\end{document}